\documentclass[aps,prl,twocolumn,floatfix,showpacs]{revtex4}

\usepackage{graphics}
\usepackage{graphicx}
\usepackage{amssymb}
\usepackage{amsfonts}
\usepackage{amsmath}
\usepackage{bm}
\usepackage{epsf}
\usepackage{version}
\usepackage[usenames]{color}

\newcommand \be  {\begin{equation}}
\newcommand \ee  {\end{equation}}
\newcommand \bea {\begin{eqnarray}}
\newcommand \eea {\end{eqnarray}}
\newcommand \nn  {\nonumber}
\newcommand \bd  {\begin{details}}
\newcommand \ed  {\end{details}}

\begin{document}

\title{Electric Excitation of Spin Resonance 
in Antiferromagnetic Conductors}
\author{Revaz Ramazashvili}
\affiliation{
Department of Physics and Astronomy, 
University of South Carolina, 
Columbia, SC 29208, USA}

\date{\today}

\excludeversion{details}


\begin{abstract}
Antiferromagnetism couples electron spin to its 
orbital motion, thus allowing excitation of electron 
spin transitions by an AC \textit{electric} 
rather than magnetic field -- with absorption, 
exceeding that of common electron spin resonance 
(ESR) at least by four orders of magnitude. 
In addition to potential applications in spin 
electronics, this phenomenon may be used as a 
spectroscopy to study antiferromagnetic materials 
of interest -- from chromium to borocarbides, cuprates, 
iron pnictides, organic and heavy fermion conductors.
\end{abstract}

\pacs{75.50.Ee,76.30.-v,76.40.+b,71.70.Ej}

\maketitle

\section{I. Introduction}

Broad research effort has been underway 
\cite{awschalom,maekawa_1,zutic} 
to build a new generation of electronic devices, 
that would manipulate and monitor carrier spin 
and charge on an equal footing. Magnetic semiconductors 
\cite{mettis,
jungwirth} 
and giant magnetoresistance materials \cite{maekawa_2}, 
as well as semiconductors with spin-orbit interaction 
\cite{rashba_SOI_2}, have been much scrutinized with 
this goal in mind.

By contrast, antiferromagnets have enjoyed 
far less attention in this context. Here, 
I show that, in fact, antiferromagnets in 
their ordered state may prove useful for 
spin ma\-ni\-pu\-lation by electric field, 
as antiferromagnetism couples electron spin 
to its orbital motion. This coupling 
ma\-ni\-fests itself especially vividly 
in a magnetic field, where it takes the 
form of anisotropic Zeeman interaction 
with a momentum-dependent $g$-tensor. This 
dependence turns a common Zeeman term into 
a spin-orbit coupling $\mathcal{H}_{ZSO}$: 
\be 
\label{eq:ZSO_coupling}
\mathcal{H}_{ZSO}
 = 
 - \mu_B 
\left[ 
g_\| 
({\bf H_\| \cdot {\bm \sigma}}) 
 +
g_\perp ({\bf p}) 
({\bf H_\perp \cdot {\bm \sigma}})
\right].
\ee
Hereafter, 
$
{\bf H}_\| 
 =
({\bf H} \cdot {\bf n})
 {\bf n}
$  
and
$
{\bf H}_\perp
 =
 {\bf H}
 - 
{\bf H}_\| 
$ 
are the longitudinal and transverse 
components of the magnetic field ${\bf H}$ 
with respect to the unit vector ${\bf n}$ 
of the staggered magnetisation, 
$
\mu_B 
$ 
is the Bohr magneton, while 
$g_\|$ 
and $g_\perp ({\bf p})$ are the longitudinal and 
transverse components of the $g$-tensor. While  
$g_\|$ is momentum-independent up to small 
relativistic corrections, $g_\perp ({\bf p})$ 
has a set of zeros in the Brillouin zone and 
thus substantially depends on the quasiparticle 
momentum momentum ${\bf p}$. 
This remarkable fact is dictated by 
the symmetry of antiferromagnetic state 
\cite{braluk,symshort,symlong}, 
and gives rise to a number of interesting effects.

One such effect amounts to excitation of spin 
resonance transitions by an AC {\em electric} 
field, with resonance absorption exceeding 
that of common electron spin resonance (ESR) 
by over four orders of magnitude. 
This phenomenon does not rely on the presence 
of localized magnetic moments, and is possible 
both for itinerant electrons and for impurity-bound 
electron states. Hence it can be used as a resonance  
spectroscopy, tailor-made to study antiferromagnets 
of great interest from chromium to cuprates, 
borocarbides, iron pnictides, as well 
as organic and heavy-fermion materials.

\begin{figure}[h]
 \hspace{3cm}
 \epsfxsize=8cm
 \epsfbox{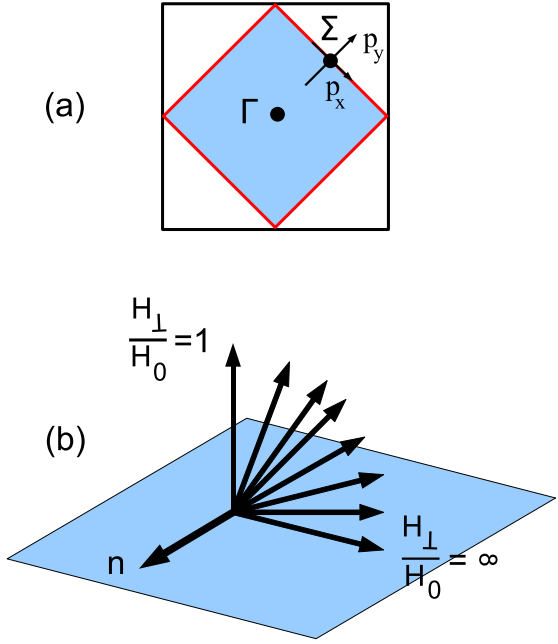}
 \vspace{15pt}
\caption{ (color online). Geometry of the problem. 
(a) The Brillouin zone of a N\'eel antiferromagnet 
on a lattice of square symmetry, and its magnetic 
Brillouin zone (MBZ, shaded diagonal square). 
The line of zero $g_\perp ({\bf p})$ must contain 
the entire MBZ boundary, shown in red online.  
Point $\Gamma$ is the Brillouin zone center, 
point $\Sigma$ is the center of the MBZ boundary, 
where the conduction band minimum is assumed to occur, 
and $p_y$ is the component of the momentum deviation from 
the minimum, locally transverse to the MBZ boundary. 
(b) Real-space geometry: staggered magnetisation 
axis ${\bf n}$, pointing along the conduction plane, 
and nearly transverse magnetic field ${\bf H}$, 
here drawn normal to ${\bf n}$; components $H_\perp$ and 
$H_0$ are normal to ${\bf n}$ and to the conducting plane, 
respectively. 
} 
\label{fig:figure_1}
\end{figure}

\section{II. The spectrum}

Here, I illustrate this effect by an example, 
that may be relevant to a number of 
antiferromagnetic conductors: I study electric 
excitation of \textit{itinerant}-electron 
resonance in a weakly-doped two-dimensional 
antiferromagnetic insulator on a lattice of square 
symmetry, whose conduction band minimum falls at the 
center $\Sigma$ of the magnetic Brillouin zone (MBZ) 
boundary, as shown in Fig. \ref{fig:figure_1}(a). 
Both the two-dimensionality and the square symmetry 
of this example simplify the description and make 
it relevant to materials such as cuprates and iron 
pnictides, yet neither of the two features 
is essential to the effect. Numerous other 
antiferromagnets of different crystal symmetry and 
effective di\-men\-sio\-na\-li\-ty are discussed in 
Ref. \cite{symlong}. Magnetic field is assumed small 
on the scale of the electron excitation gap $\Delta$ 
and of the reorientation threshold, and thus does 
not perturb antiferromagnetic order.

The effect is most vivid for the staggered 
magnetization axis $\bf{n}$, pointing along the 
conducting plane, which is the case in several 
electron-doped cuprates \cite{lavrov,matsuda}. 
The magnetic field ${\bf H}$ is nearly normal 
to ${\bf n}$, which tends to happen due to 
spin-flop. It is this very geometry that 
I consider hereafter; orientation of the field 
with respect to the conducting plane may be 
arbitrary, as shown in Fig. \ref{fig:figure_1}(b). 

At low doping, the carriers concentrate in a small 
vi\-ci\-ni\-ty of the band minimum $\Sigma$, 
and the Hamiltonian can be expanded around it.     
By symmetry, $g_\perp ({\bf p})$ in $\mathcal{H}_{ZSO}$ 
(\ref{eq:ZSO_coupling}) 
vanishes upon approaching the MBZ boundary, linearly 
in a generic case \cite{symshort,symlong} -- and can 
be recast as 
$g_\perp({\bf p}) = g_\| \frac{p_y \xi}{\hbar}$ 
with a constant $\xi$, for $\frac{p_y \xi}{\hbar} \ll 1$. 
Here, $p_y$ is the component of the momentum deviation 
from the band minimum, locally transverse to the 
MBZ boundary, as shown in Fig. \ref{fig:figure_1}(a). 
The length scale $\xi$ is of the order of the 
antiferromagnetic coherence length $\hbar v_F/\Delta$, 
and may be of the order of the lattice constant 
or much greater \cite{symshort,symlong}. 

Near the band minumum, the kinetic energy is quadratic 
in ${\bf p}$; 
for simplicity, I consider isotropic effective mass $m$, 
and introduce ${\bf \Omega} \equiv \mu_B g_\| {\bf H}$. 
For the field at a finite angle with the 
conducting plane, the Hamiltonian reads 
\begin{equation}
\label{eq:Hamiltonian_point_X}
\mathcal{H} = 
\frac{1}{2m} 
\left[
{\bf p} - \frac{e}{c}{\bf A}
\right]^2
 - 
({\bf \Omega}_\| \cdot {\bm \sigma}) 
 - 
\frac{\xi}{\hbar} 
\left[
{\bf p} - \frac{e}{c}{\bf A}
\right]_y
 ({\bf \Omega}_\perp \cdot {\bm \sigma})
,
\end{equation}
where ${\bf A}$ 
is the electromagnetic vector potential 
\cite{parity}.

In a purely transverse field (${\bf \Omega}_\|=0$), 
the up- and the down-spin projections onto 
${\bf \Omega}_\perp$ decouple and have identical 
Landau spectra: 
\be
\label{eq:quantized_spectrum_X}
\mathcal{E}_{n}
 = 
\Omega_0
\left[ 
n + \frac{1}{2} 
\right], 
\ee 
where 
$
\Omega_0 \equiv \hbar \frac{e H_0}{m c}
$ 
is the cyclotron energy, and $H_0$ 
is the normal component of the field 
with respect to the conducting plane. 
This degeneracy becomes explicit upon 
completing the square in Eqn. 
(\ref{eq:Hamiltonian_point_X})
with respect to 
$[{\bf p} - \frac{e}{c}{\bf A}]_y$, or upon 
performing a non-uniform spin rotation 
\be
\label{eq:symmetry_1}
\Psi \rightarrow 
\exp 
\left[
i 
\frac{y}{\hbar} 
\frac{m \xi 
}{\hbar} 
({\bf \Omega}_\perp \cdot {\bm \sigma})
\right]
\Psi,
\ee
which, in a purely transverse field 
(${\bf \Omega}_\| = 0$),  
eliminates ${\bf \Omega}_\perp$ from the Hamiltonian altogether. 

\begin{figure}[h]
 \hspace{3cm}
 \epsfxsize=8cm
 \epsfbox{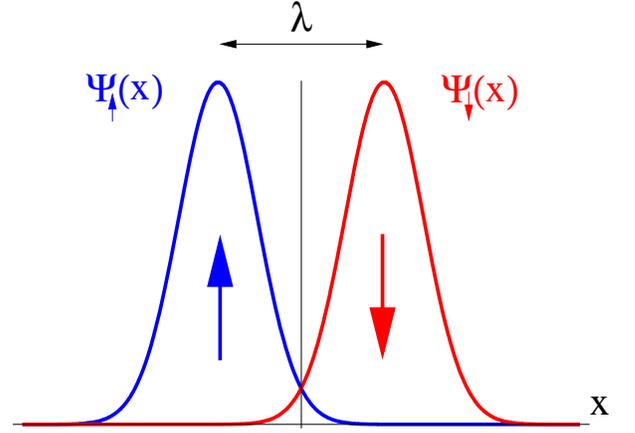}
 \vspace{15pt}
\caption{(color online). 
Splitting of degenerate spin states in real space: 
the spin `up' state $\Psi_\uparrow (x)$ and the spin 
`down' state $\Psi_\downarrow (x)$ at the lowest 
Landau level, with the spin quantization axis chosen 
along ${\bf \Omega}_\perp$. In a purely transverse 
field \mbox{(${\bf \Omega}_\| = 0$)}, the two wave 
functions remain degenerate, but split by the distance 
\mbox{$\lambda = 2 \xi \frac{\Omega_\perp}{\Omega_0}$} 
along the $\hat{x}$ axis in real space. 
} 
\label{fig:real_space_splitting}
\end{figure}

In the Landau gauge \mbox{${\bf A} = (0, x H_0)$}, 
this spin de\-ge\-ne\-ra\-cy in a transverse field 
acquires a simple interpretation: as shown in Fig. 
\ref{fig:real_space_splitting}, the guiding orbit 
centers of the spin-up and the spin-down states 
split apart by the distance 
$
\lambda 
  \equiv 
 2 \xi 
\frac{
\Omega_\perp}{
\Omega_0} 
$ along the $\hat{x}$ axis in real space, 
with the spin quantization axis chosen along 
${\bf \Omega}_\perp$.

To study the spectrum in an arbitrary field, 
it is convenient to use a different Landau gauge:   
${\bf A} = (-y H_0, 0)$. 
The spin rotation (\ref{eq:symmetry_1}) removes 
the transverse field term, and turns the uniform 
longitudinal field ${\bf \Omega}_\|$ 
into a spiral texture ${\bf \Omega}_\|'$  
with a constant pitch $q \equiv 
\frac{2 m \xi \Omega_\perp}{\hbar^2}$ 
along the $\hat{y}$-axis in the conducting plane:
\be
{\bf \Omega}_\|'
   =  
\label{eq:texture}
{\bf \Omega}_\|
\cos 
 \left[q y
   \right]
  + 
 {\bf n}_\perp
  \times 
{\bf \Omega}_\|
\sin
\left[ 
q y 
\right], 
\ee
where ${\bf n}_\perp$ is the unit vector 
along ${\bf \Omega_\perp}$. It is helpful 
to recast the cyclotron motion in terms of 
ladder operators as per 
$
\frac{a + a^+}{\sqrt2} 
 \equiv 
\frac{y}{l_H} - 
\frac{p_x l_H}{\hbar}
$ 
and  
$\frac{a - a^+}{i \sqrt2} \equiv 
\frac{p_y l_H}{\hbar}$, 
where $l_H = \sqrt\frac{\hbar c}{e H_0}$ 
is the magnetic length. 
Now, the Hamiltonian 
(\ref{eq:Hamiltonian_point_X}) reads 
\be
\mathcal{H} 
  =  
\label{eq:Hamiltonian_a+a_transformed}
\Omega_0 
\left[ 
a^+ a 
  + 
\frac{1}{2} 
\right]
 - 
({\bf \Omega}_\|' \cdot {\bm \sigma}), 
\ee
with $y$ in ${\bf \Omega}_\|'$ 
of Eqn. (\ref{eq:texture}) 
expressed via the ladder 
o\-pe\-ra\-tors.

\begin{figure}[h]
 \hspace{3cm}
 \epsfxsize=8cm
 \epsfbox{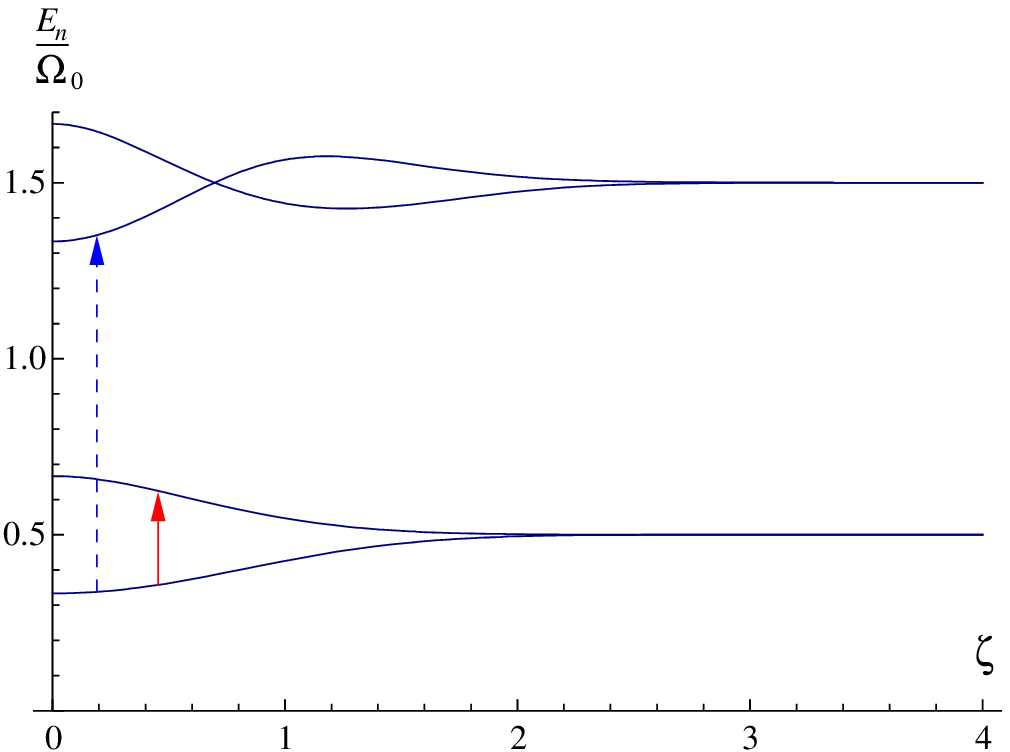}
 \vspace{15pt}
\caption{(Color online) The Landau level splitting. 
The first two Landau levels $E_n$, split by the 
longitudinal field \mbox{$\Omega_\| = \Omega_0/6$}, 
are shown in units of $\Omega_0$ as a function of 
\mbox{$
\zeta = 
\frac{\xi}{l_H} 
\frac{\Omega_\perp}{\Omega_0}
$}. 
The levels were obtained by numerical solution of 
the Hamiltonian (\ref{eq:Hamiltonian_a+a_transformed}), 
truncated to the lowest six levels. The solid arrow 
indicates the ZEDR transition at the lowest Landau 
level, whereas the dashed arrow corresponds to the 
cyclotron resonance transition from the lowest 
to the first Landau level. 
} 
\label{fig:figure_3}
\end{figure}

According to Eqn. 
(\ref{eq:Hamiltonian_a+a_transformed}), 
in the limit of a weak longitudinal field 
($\Omega_\| \ll \Omega_0$), the spin precesses 
at a characteristic frequency $\Omega_\|$, 
which is small compared with the cyclotron 
frequency $\Omega_0$ of the orbital motion. 
In this limit, the splitting 
$\delta \mathcal{E}_{n}$ of the $n$-th 
Landau level is given simply by averaging 
$({\bf \Omega}_\|' \cdot {\bm \sigma})$ 
over the orbital eigenstate $| n \rangle$ 
of the first term in Eqn. 
(\ref{eq:Hamiltonian_a+a_transformed}), 
leading to 
\be
\label{eq:LL_splitting_X}
\delta \mathcal{E}_{n} =
2 
\Omega_\| 
f_n
\left(
\frac{\xi}{l_H} 
\frac{\Omega_\perp}{\Omega_0}
\right), 
\ee
where 
$
f_n(\zeta)
 =
L_n (2\zeta^2)
 \exp[-\zeta^2] 
$,
and 
$L_n(\zeta)$ 
is the Laguerre polynomial \cite{grad1}. 
The degeneracy is lifted in a peculiar way: 
for $\Omega_\| \ll \Omega_0$, the splitting 
$\delta \mathcal{E}_{n}$ of the $n$-th Landau 
level decays and oscillates as a function of 
$\zeta = \frac{\xi}{l_H} 
\frac{\Omega_\perp}{\Omega_0}$, 
as shown in Fig. \ref{fig:figure_3}. 
For a small fixed $\Omega_\| \ll \Omega_\perp$, 
this amounts to decaying oscillations with reducing 
the angle between the field and the conducting plane,  
as shown in Fig. \ref{fig:figure_1}(b).

The factor $f_n(\zeta)$ in Eqn. 
(\ref{eq:LL_splitting_X}) is of a simple origin. 
The longitudinal component ${\bf \Omega}_\|$ 
hybridizes the two states in Fig. 
\ref{fig:real_space_splitting} and lifts their 
degeneracy. Indeed, the splitting vanishes 
as the distance 
$\lambda = 2 \xi \frac{\Omega_\perp}{\Omega_0}$ 
between the guiding orbit centers exceeds 
the wave function size $l_H \sqrt{n + 1}$. 
The oscillations on the background of this 
decay are due to spatial oscillation of 
the two wavefunctions for $n > 0$. 

\section{III. Electric excitation of spin resonance}

The momentum dependence of $g_\perp ({\bf p})$ 
has a spec\-ta\-cular spectroscopic manifestation: 
excitation of spin resonance transitions by an AC 
{\em electric} field -- the very same transitions 
that are normally excited by an AC magnetic field 
in an ESR experiment. 

I name this phenomenon Zeeman Electric-Dipole 
Resonance (ZEDR) to note its similarity with 
Electric-Dipole Spin Resonance (EDSR) 
in semiconductors and semiconducting 
heterostructures with spin-orbit coupling 
\cite{rashba_book}. 

\subsection{A. Resonance in a quantizing field}

To study the effect for discrete Landau levels, 
notice that a uniform AC electric field $E^y_\omega$ 
along the $\hat{y}$-axis couples to the $y$-component 
$ey = e l_H \frac{a + a^+}{\sqrt2}$ of the electron 
dipole moment. With $E^y_\omega$, the Hamiltonian  
(\ref{eq:Hamiltonian_a+a_transformed}) reads  
\be
\label{eq:Hamiltonian_a+a_transformed_t-dependent}
\mathcal{H} 
  = 
\Omega_0 
\left[ 
a^+ a 
 + 
\frac{1}{2}
\right]
- 
({\bf \Omega}_\|' \cdot {\bm \sigma}) 
- e 
\frac{a + a^+}{\sqrt2} 
l_H E^y_\omega.
\ee

In the absence of a longitudinal component 
${\bf \Omega}_\|$, the last term in 
(\ref{eq:Hamiltonian_a+a_transformed_t-dependent}) 
induces only the cyclotron resonance: spin-conserving 
electric dipole transitions between the adjacent 
Landau levels, with the matrix element $M_{CR}$ 
\be
\label{eq:CR_matrix_element_quantised}
M_{CR} 
 =  
\langle 
n+1, \sigma 
 | 
e y E^y_\omega
 | 
 n, \sigma 
 \rangle 
 = 
 e l_H E_\omega^y 
\sqrt{
\frac{n + 1}{2}
},  
\ee
whose scale is set by the 
Larmore radius $l_H \sqrt{n + 1}$. 

A small longitudinal component 
$\Omega_\| \ll \Omega_0$ 
changes this picture, as 
$({\bf \Omega}_\|' \cdot {\bm \sigma})$ 
couples the electron spin to its orbital motion. 
As a result, the $n$-th Landau level eigenstate 
$ | n \alpha \rangle$ with spin projection 
$\alpha$ on the direction of ${\bf \Omega}_\|$
acquires a small admixture of other states 
$ | m \beta \rangle$, and the AC electric field 
begins to induce a number of previously forbidden 
transitions.

Here, I restrict myself to spin-flip transitions 
within the same Landau level \cite{zedr}, excited 
by an AC {\em electric} field as shown in Fig. 
(\ref{fig:figure_3}). Treating the admixture 
of other Landau levels to the first order in 
$({\bf \Omega}_\|' \cdot {\bm \sigma})$, 
one finds \cite{revaz_erratum} the ZEDR 
matrix element 
$
M_{ZEDR} \equiv
\langle n \uparrow 
 |
e y E_\omega^y 
 |
 n \downarrow 
\rangle 
$: 
\be
\label{eq:ZEDR_matrix_element_quantised} 
M_{ZEDR} 
 = 
  - 2 e \xi E_\omega^y 
\frac{\Omega_\|}{\Omega_0} 
\frac{\Omega_\perp}{\Omega_0}
L_n 
(2 \zeta^2)
\exp
\left[
- \zeta^2
\right], 
\ee
where
$\zeta \equiv 
\frac{\xi}{l_H} 
\frac{\Omega_\perp}{\Omega_0}
$. 
Apart from the dependence on the orientation of 
the field with respect to the conducting plane and to 
the staggered magnetization, ZEDR matrix elements are 
defined simply by the length scale $\xi$. 
Being at least of the order of the lattice spacing, 
in a weakly-coupled spin density wave antiferromagnet 
$\xi \sim \hbar v_F/\Delta$ 
(see Refs. \cite{symshort,symlong})  
may reach a $10$ nm scale \cite{CR}. 
At the same time, the ESR matrix elements are defined 
by the Compton length 
$\lambda_C = \frac{\hbar}{mc} \approx 0.4$ pm. 
The characteristic ratio of the ZEDR matrix 
elements to those of ESR can thus be estimated as 
$\frac{\xi}{\lambda_C}
 = \frac{1}{\alpha} \frac{\xi}{a_B}$, 
where $a_B = \frac{\hbar^2}{m e^2} \approx 53$ pm 
is the Bohr radius, and 
$\frac{1}{\alpha} = \frac{\hbar c}{e^2} \approx 137$ 
is the inverse fine structure constant. 
Thus, the ZEDR absorption exceeds that of ESR by 
about $\left[ 137 \cdot \frac{\xi}{a_B} \right]^2$, 
which amounts to at least four orders of magnitude. 

\subsection{B. Resonance in a continuous spectrum}

Now, consider a situation, where the DC magnetic field 
${\bf H}$ couples only to the electron spin, but not 
to its orbital motion, which is the case for a field 
along the conducting plane. According to Eqn. 
(\ref{eq:Hamiltonian_point_X}),  
the field splits the conduction band into 
two subbands $\mathcal{E}_\pm ({\bf p})$ 
\begin{equation}
\label{eq:continuous_splitting_X}
\mathcal{E}_\pm ({\bf p}) =
\frac{{\bf p}^2}{2m} \pm
\sqrt{ \Omega_\|^2 + 
\left[ 
\frac{p_y \xi}{\hbar} 
\right]^2 
\Omega_\perp^2 }
,
\end{equation}
and the AC field induces 
transitions between them. 

\begin{details}
Notice that the momentum dependence of 
$g_\perp ({\bf p})$ renders the direction 
$(
{\bf \Omega}_\| + 
\frac{\xi p_y}{\hbar} 
{\bf H_\perp})$  
of the preferential spin quantization 
axis momentum-dependent. 
\end{details}
\begin{figure}[h]
 \hspace{3cm}
 \epsfxsize=6cm
 \epsfbox{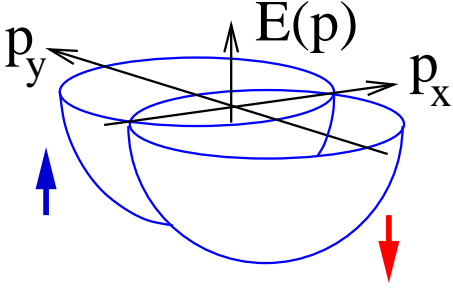}
 \vspace{15pt}
\caption{(Color online) The spin splitting 
of the conduction band, sketched after Eqn. 
(\ref{eq:Hamiltonian_point_X_contimuous}) 
in a small vicinity of the band mi\-ni\-mum 
at point $\Sigma$.
} 
\label{fig:figure_4}
\end{figure}

According to (\ref{eq:Hamiltonian_point_X}), 
a purely transverse field ($\Omega_\|=0$) 
lifts the Kramers degeneracy by splitting the 
two de\-ge\-ne\-rate subbands by the `distance' 
\mbox{$\delta p_y = \frac{2}{\hbar} m \xi \Omega_\perp$} 
along the $p_y$-axis:  
\begin{equation}
\label{eq:Hamiltonian_point_X_contimuous}
\mathcal{H} = 
\frac{p_x^2}{2m}
 + 
\frac{1}{2m} 
\left[
p_y - \frac{m \xi}{\hbar} 
({\bf \Omega}_\perp \cdot {\bm \sigma})
\right]^2. 
\end{equation} 
Illustrated in Fig. \ref{fig:figure_4},  
this is, indeed, a momentum-space counterpart 
of the real-space splitting in Fig.  
\ref{fig:real_space_splitting}.

In the continuous spectrum, ZEDR may be treated 
simply as being induced by the term 
\mbox{
$
\delta\mathcal{H}_{ZEDR}^\omega 
 = \frac{\xi}{\hbar} \frac{e}{c} 
A_\omega^y 
({\bf \Omega}_\perp \cdot \sigma) 
$}.  
Its matrix element between the states with 
the spin along and against the direction 
of the effective magnetic field 
\mbox{
$(
{\bf \Omega}_\|+
\frac{\xi p_y }{\hbar}
{\bf \Omega}_\perp)$} 
is equal to 
\begin{equation}
\label{eq:matrix_element_continuous_case}
|\langle \uparrow |
 \delta\mathcal{H}_{ZEDR}^\omega
 | \downarrow \rangle|^2 = 
\left[ 
\frac{e \xi E^y_\omega}{\hbar \omega} 
\right]^2 
\frac{\Omega_\|^2 \Omega_\perp^2}
{
\Omega_\|^2
 + 
\left[ 
\frac{\xi p_y}{\hbar} 
\Omega_\perp \right]^2
}, 
\end{equation}
where I used the relation     
$\langle\downarrow|(\hat{n}\cdot\sigma)|\uparrow\rangle
= n_+ \equiv n_x + in_y$ for an arbitrary unit vector $\hat{n}$.

The ZEDR absorption $P_{ZEDR}^\omega$ 
is given, according to the Fermi golden 
rule, by the product of the modulo squared 
(\ref{eq:matrix_element_continuous_case}) 
of the matrix element of $\delta\mathcal{H}_{ZEDR}^\omega$ 
by the AC field frequency $\omega$, 
and by $\frac{\pi}{\hbar}$, with the 
subsequent summation over the Fermi 
surface, yielding
\begin{details}
\be
\label{eq:absorption_X}
P_{CR}^\omega
 = 
\frac{[e \xi E_\omega^y H_\perp H_{||}]^2}{\omega^3} 
\frac{4\pi}{\hbar}
\sum_{\bf p} 
\delta 
\left( 
\omega - 2 
\sqrt{ H_{||}^2 + 
\left[\frac{p_y \xi}{\hbar}\right]^2 H_\perp^2 }
\right)
\delta(\eta({\bf p})-\mu)
.
\ee
Up to the factor of $g_{||}$ (set to unity 
in \cite{revaz}), and to the normalisation 
factor, this corresponds to formula (3) 
in \cite{revaz}.
Summation over ${\bf p}$ yields 
\end{details}
\begin{widetext}
\be
\label{eq:absorption_X_answer}
P_{ZEDR}^\omega = 
\frac{
m
}{\pi}
\frac{[e \xi E_\omega^y]^2}{16} 
\frac{
  \sin^2 \theta \cos^2 \theta 
\left(
\frac{\omega_H}{\hbar \omega}
\right)^4
}
{
\sqrt{
\left(
\left[
 \frac{
\omega_H}{\hbar \omega}
\right]^2
\left[
\cos^2 \theta + 2\mu m \xi^2
\sin^2 \theta 
\right] - 1
\right)
\left(
1
 - 
 \left[
\frac{\omega_H}{\hbar \omega}
 \right]^2 
\cos^2 \theta
\right)
  }
}, 
\ee
\end{widetext}
where $\mu$ is the electron chemical potential 
counted from the bottom of the band, and 
$\omega_H \equiv 2 \Omega$. The result \cite{revaz} 
is presented in a form, corresponding to sweeping 
the magnitude of the DC field at a fixed angle 
$\theta$ to the staggered magnetisation ${\bf n}$ 
and at a fixed frequency $\omega$. In agreement with 
Eqn. (\ref{eq:ZEDR_matrix_element_quantised}), 
the ZEDR matrix elements are again 
defined simply by the lengthscale $\xi$. 

The lineshape described by Eqn. 
(\ref{eq:absorption_X_answer}) is intrinsically broadened:  
according to Eqn.(\ref{eq:continuous_splitting_X}), in 
a magnetic field of a generic orientation, each point 
at the Fermi surface has its own resonance frequency. 
Hence the absorption is non-zero in a finite interval  
of frequencies, with square-root singularities at the 
edges. This intrinsic broadening may be a reason 
behind the ESR silence of the cuprates \cite{shengelaya}. 

\begin{details}
Integration in detail:
\bea
\nn
P_{CR}^\omega
 & = & 
\omega 
\left[ 
\frac{eE_y^\omega}{\omega} 
\right]^2 
\frac{\pi}{\hbar}
\sum_{\bf p}
[\partial_i g_\perp ({\bf p})]^2
\frac{[H_\perp g_{||}H_{||}]^2}
{[g_{||}H_{||}]^2 + [g_\perp({\bf p}) H_\perp]^2} \times \\
 & \times &
\delta 
\left[ 
\omega - 2 \sqrt{ [g_{||} H_{||}]^2 + [g_\perp ({\bf p}) H_\perp]^2 }
\right]
\delta(\eta({\bf p})-\mu) = || \,\, drop \,\, \hbar \,\, || = \\
 & = &
\nn
\frac{[e E_y^\omega H_\perp g_{||}^2 H_{||}]^2}{q_0^2 \omega^3} 
 4\pi
\sum_{\bf p} 
\delta 
\left( 
\omega - 2 g_{||} 
\sqrt{ H_{||}^2 + \left[\frac{p_y}{q_0}\right]^2 H_\perp^2 }
\right)
\delta(\eta({\bf p})-\mu) = \\
 & = & 
\nn
|| \,\, H_{||}=H\cos\theta, \,\, H_\perp = H \sin \theta  \,\, || = 
\frac{[e E_y^\omega H_\perp g_{||}^2 H_{||}]^2}{q_0^2 \omega^3} 
 4\pi \times \\
 & \times & 
\nn
\sum_{\bf p} 
\delta 
\left( 
\omega - 2 g_{||} H 
\sqrt{ \cos^2 \theta + \left[\frac{p_y}{q_0}\right]^2 \sin^2 \theta }
\right)
\delta(\frac{p_x^2}{2m_x} + \frac{p_y^2}{2m_y} - \mu) = \\
 & = & 
\nn
\frac{[e E_y^\omega H_\perp g_{||}^2 H_{||}]^2}{q_0^2 \omega^3} 
 4\pi
\sum_{\bf p} 
\delta 
\left( 
\omega - 2 g_{||} H 
\sqrt{ \cos^2 \theta + \left[\frac{p_y}{q_0}\right]^2 \sin^2 \theta }
\right) \times \\
 & \times &
\nn
\delta 
 \left[
 \frac{p_x^2}{2\sqrt{m_x m_y}} \sqrt{\frac{m_y}{m_x}}
 +
 \frac{p_y^2}{2\sqrt{m_x m_y}} \sqrt{\frac{m_x}{m_y}}
 - \mu 
\right] = \\ 
 || 
 & & 
\nn
k_x \equiv p_x \left[\frac{m_y}{m_x} \right]^{1/4}, 
k_y \equiv p_y \left[\frac{m_x}{m_y} \right]^{1/4}, 
 \,\,\,\, 
k_y = k \sin \phi = \sqrt{k_x^2 + k_y^2} \sin \phi
 || \\
 & = & 
\nn
\frac{[e E_y^\omega H_\perp g_{||}^2 H_{||}]^2}{q_0^2 \omega^3} 
 4\pi
\sum_{\bf k} 
\delta 
\left( 
\omega - 2 g_{||} H 
\sqrt{ \cos^2 \theta + \left[\frac{k}{q_0}\right]^2 
\sqrt{\frac{m_y}{m_x}} 
 \sin^2 \theta 
\sin^2 \phi 
}
\right)
 \times \\
 & \times &
\nn
\delta 
 \left[
 \frac{k_x^2 + k_y^2}{2\sqrt{m_x m_y}}
 - \mu 
\right] = 
\frac{[e E_y^\omega H_\perp g_{||}^2 H_{||}]^2}{q_0^2 \omega^3} 
\int 
\frac{kdkd\phi}{\pi}
\delta 
 \left[
 \frac{k^2}{2\sqrt{m_x m_y}}
 - \mu 
\right] \times \\
 & \times & 
\nn 
\delta 
\left( 
\omega - 2 g_{||} H 
\sqrt{ \cos^2 \theta + \left[\frac{k}{q_0}\right]^2 
\sqrt{\frac{m_y}{m_x}} 
 \sin^2 \theta 
\sin^2 \phi 
}
\right) = 
\frac{[e E_y^\omega H_\perp g_{||}^2 H_{||}]^2}{q_0^2 \omega^3} 
 \times
\\
 & \times & 
\nn
\sqrt{m_x m_y}
\int 
\frac{d\phi}{\pi}
\delta 
\left[
\omega - 2 g_{||} H 
\sqrt{ \cos^2 \theta + \frac{2 \mu m_y}{q_0^2}
 \sin^2 \theta 
\sin^2 \phi 
}
\right] = 
\eea
\bea
 & & 
\nn
 || \,\, 
\int d\alpha \delta
\left[
a - \sqrt{b^2 + c^2 \sin^2 \alpha}  
\right] = 
\frac{a}{\sqrt{a^2 - b^2}\sqrt{b^2 + c^2 - a^2}} \,\, ||
 \\
 & & 
\nn
 || \,\, 
\omega_{\min} \equiv 2g_{||}H_{||}, \,\,
\omega_{\max} \equiv 2g_{||} 
\sqrt{H_{||}^2 + \frac{2 \mu m_y}{q_0^2} H_{\perp}^2} \,\, || 
 \\
 & = & 
\nn
\frac{[e E_y^\omega H_\perp g_{||}^2 H_{||}]^2}{q_0^2 \omega^3} 
\frac{\sqrt{m_x m_y}}{\pi}
\frac{\omega}
{
\sqrt{
\omega_{\max}^2
 - \omega^2
}
\sqrt{\omega^2 - \omega_{\min}^2}
}
 =
 \\
 & = & 
\nn
\frac{\sqrt{m_x m_y}}{\pi}
\frac{[e E_y^\omega]^2}{16 q_0^2} 
\frac{
  \sin^2 \theta \cos^2 \theta 
\left(
\frac{2g_{||}H}{\omega}
\right)^4
}
{
\sqrt{
\left(
 \frac{2g_{||}H}{\omega}
\right)^2
\left[
\cos^2 \theta + \frac{2\mu m_y}{q_0^2}
\sin^2 \theta 
\right] - 1
}
\sqrt{ 1
 - 
 \left(
\frac{2g_{||}H}{\omega}
 \right)^2 
\cos^2 \theta
  }
}
\eea
\end{details}

\section{IV. Discussion and conclusions}

Electric excitation of spin resonance becomes 
possible due to a substantial variation of the 
$g$-tensor across the Brillouin zone. 
In antiferromagnetic conductors, this variation 
is imposed by symmetry \cite{braluk,symshort,symlong}. 
Hence, ZEDR shall be found in a broad range 
of materials from weakly-doped antiferromagnetic 
insulators to antiferromagnetic metals. Quantitative 
details between these two limits may vary, 
but the key sufficient condition for ZEDR amounts 
to a significant variation of $g_\perp ({\bf p})$ 
for the actual carriers.

Zeeman Electric-Dipole Resonance is induced by an 
AC electric field; to study it, a small sample has 
to be placed in a resonator at the electric field 
maximum. This puts ZEDR in competition against 
cyclotron resonance (CR), the latter generally being 
a stronger effect. Nevertheless, these two resonances 
can be easily distinguished. 

Firstly, the CR and the ZEDR frequencies are different.  
In a nearly transverse field ($\Omega_\| \ll \Omega_0$), 
the former is simply the cyclotron frequency $\Omega_0$ 
up to small corrections of the order of 
$\Omega_\|/\Omega_0 \ll 1$, 
as shown in Fig. \ref{fig:figure_3}. 
The ZEDR frequency is much smaller, of the 
order of $\Omega_\| \ll \Omega_0$, and shows 
the peculiar dependence (\ref{eq:LL_splitting_X}) 
on the magnetic field strength and orientation. 

Secondly, the ZEDR absorption grows with increasing 
the magnetic field and, already in a low field 
$\Omega_0 \sim \frac{\Delta^2}{\epsilon_F}$, 
becomes of the same order of magnitude as that 
of cyclotron resonance: this follows from Eqns. 
(\ref{eq:CR_matrix_element_quantised}) and  (\ref{eq:ZEDR_matrix_element_quantised}) 
\cite{expansion_limit}. 
For materials with $\Delta \ll \epsilon_F$, 
this crossover scale is small compared with $\Delta$, 
which means that the ZEDR intensity may exceed that 
of the cyclotron resonance while the field is still 
much smaller than $\Delta$, and hence does not 
perturb the antiferromagnetic order. This makes 
an antiferromagnetic conductor with a small ratio  
$\frac{\Delta}{\epsilon_F} \ll 1$ a promising 
candidate for the observation of ZEDR.

This condition is met by a number of materials 
from weakly-doped antiferromagnetic insulators 
to antiferromagnetic metals with a large Fermi 
surface. Among the latter, the simplest of 
opportunities to observe ZEDR may be offered 
by chromium \cite{fawcett_1}, an archetypal 
spin density wave metal, ever attracting 
much attention \cite{kummamuru}. 

Among systems more complex, oxychlorides  
\cite{ronning} and electron-doped cuprates  
\cite{armitage} have recently shown the 
appearance of carriers near the point $\Sigma$ 
at the MBZ boundary at low doping, which may 
allow ZEDR, provided that the antiferromagnetic 
correlations are de\-ve\-loped well enough. 
A number of other relevant materials 
are discussed in \cite{symlong}. 

\subsection{A. Experimental issues}

In this subsection, I discuss a number of issues,  
that may be important for a successful observation 
of electrically-excited electron spin resonance in 
an antiferromagnetic conductor. 

Since the effect hinges on a substantial momentum 
dependence of the $g$-tensor, it requires clean 
samples. Observation of de Haas--van Alphen  
oscillations could serve as an experimental 
criterion of a sufficient sample purity. Similarly, 
it is desirable to work with single-magnetic-domain 
samples.

Thermal fluctuations of the antiferromagnetic 
order reduce the ordered magnetization and 
scatter the charge carriers; at the same time, 
directional fluctuations of the staggered 
magnetization make the resonance frequency 
scale $\Omega_\|$ vary in space, leading 
to an additional smearing of the resonance. 
These effects can be suppressed by working 
well below the N\'eel temperature. 

The theory above implicitly assumed, that the 
orientation of the field ${\bf H}$ with respect 
to the staggered magnetization ${\bf n}$ and 
to the conducting plane may be varied at will. 
This requires a sufficient magnetic anisotropy 
to maintain the orientation of ${\bf n}$ with 
respect to the crystal axes, or otherwise 
spin-flop would re-orient ${\bf n}$ 
transversely to ${\bf H}$. Therefore, ${\bf H}$ 
must be kept below the reorientation field 
of the material. 

At the same time, the magnetic anisotropy helps 
to separate the electron spin resonance frequency 
from that of the antiferromagnetic resonance 
\cite{keffer}: the former scales with $\Omega_\|$, 
while the latter scales as the square root of the 
anisotropy and thus remains finite at zero field. 
Possible interference with antiferromagnetic 
resonance is conveniently suppressed even further 
due to the fact that antiferromagnetic resonance 
is excited by an AC \textit{magnetic} field, 
which has a node at an AC electric field maximum 
of the resonator.

Decay of the AC electric field beyond a thin 
surface layer of the sample (the skin effect) 
presents another challenge for the electric 
excitation of spin resonance. This issue may be 
bypassed by working with films \cite{zabel} thinner 
than the skin depth. In the relevant frequency 
range of 10 GHz, the skin depth of a good metal 
such as chromium is of the order of one $\mu$m. 
Lowering the carrier  concentration increases 
the skin depth: for various organic conductors 
and underdoped cuprates in the 10 GHz range, 
the skin depth measures dozens of $\mu$m 
\cite{oshima,trunin}. 

Previous studies of EDSR in semiconductors 
(see Refs. \cite{rashba_book,shekhter,duckheim}) 
focused on relativistic spin-orbit interaction 
terms, those that appear in the absence of 
magnetic field. By contrast, in an antiferromagnet,  
the Zeeman spin-orbit coupling $\mathcal{H}_{ZSO}$ 
of Eqn. (\ref{eq:ZSO_coupling}) is proportional 
to the magnetic field, which renders ZEDR tunable. 
Yet, as shown above, ZEDR becomes strong 
already in a weak field, which turns it 
into a promising experimental tool. 

ZEDR offers a new method to investigate the coupling 
between electron spin and its orbital motion in 
antiferromagnets. Materials from chromium to 
borocarbides, cuprates, iron pnictides, organic 
and heavy fermion compounds may be studied using 
this phenomenon, and perhaps employed to manipulate 
and monitor carrier spin with electric field. 
 
I thank I. Aleiner, B. Altshuler, S. Brazovskii 
and L. P. Pitaevskii for enlightening discussions, 
and S. Carr, Ya. Bazaliy, N. Shannon, A. Shengelaya, 
T. Vekua, J. Wykhoff and S. Zvyagin for helpful 
comments. Parts of this work were done at the 
Max-Planck Institute for the Physics of Complex 
Systems in Dresden, at LPTMS in Orsay, and at 
the Institut Henri Poincare-Centre Emile Borel 
in Paris. I thank LPTMS, MPI PKS and IHP 
for the kind hospitality, and IFRAF for its 
generous support.


\begin{thebibliography}{99}
\bibitem{parity}
At first sight, the longitudinal and the transverse 
term in Eqn. (\ref{eq:Hamiltonian_point_X}) transform  
differently under inversion $\mathcal{I}$. 
However, $\xi$ also has odd parity 
(see Refs. \cite{symshort,symlong}), 
thus removing the issue.
\bibitem{zedr}
In a weak field, transitions to other Landau 
levels are suppressed by even powers of 
$\left[\frac{\xi}{l_H}\right]^{2} \ll 1$.
\bibitem{revaz_erratum} 
ZEDR has been predicted in Ref. \cite{revaz}. 
However, the result for absorption in a quantizing 
field, given in this article, overlooks the fact, 
that both the Landau level splitting and the 
absorption vanish in a purely transverse field.
\bibitem{CR} 
For comparison, cyclotron resonance matrix elements 
in a magnetic field of one Tesla are defined 
by $l_H \approx 26$ nm.
\end{thebibliography}
\end{document}